\documentclass[sigconf]{acmart}
\usepackage[shortlabels]{enumitem}
\usepackage{CJKutf8}
\usepackage{xcolor}
\usepackage{hyperref}
\usepackage[normalem]{ulem}
\usepackage{enumitem}
\usepackage{listings}
\usepackage{array}
\usepackage{tabularx}
\usepackage{booktabs} 
\usepackage{tabularx}
\usepackage{booktabs}
\usepackage{float}
\usepackage{tcolorbox}
\newcommand{\equalcontrib}{Both authors contributed equally to this research.}
\usepackage[linesnumbered,ruled,vlined]{algorithm2e}

\hypersetup{
    colorlinks=true,
    linkcolor=blue,
    filecolor=magenta,      
    urlcolor=cyan,
    pdfborderstyle={/S/U/W 1}
}

\makeatletter
\begingroup
  \catcode`\$=6 %
  \catcode`\#=12 %
\endgroup

\def\BibTeX{{\rm B\kern-.05em{\sc i\kern-.025em b}\kern-.08emT\kern-.1667em\lower.7ex\hbox{E}\kern-.125emX}}
    
\setcopyright{rightsretained}
\copyrightyear{2025}
\acmYear{2025}
\acmConference[Genaiecom '25]{Proceedings of the second workshop on Generative AI for E-Commerce}{September 22, 2025}{Prague, CZ}
\acmBooktitle{Proceedings of the second workshop on Generative AI for E-Commerce 2025, September 22, 2025}
\acmYear{2025}
\acmMonth{09}
\acmDOI{}
\acmISBN{}
\acmPrice{15.00}

\begin{document}
%
\title{Mind the Gap: Bridging Behavioral Silos with LLMs in Multi-Vertical Recommendations}


\author{Nimesh Sinha}
\authornote{\equalcontrib}
\email{nimesh.sinha@doordash.com}
\affiliation{%
  \institution{DoorDash Inc.}
  \streetaddress{303 2nd Street, Suite 800}
  \city{San Francisco}
  \state{California}
  \postcode{94107}
  \country{USA}
}

\author{Raghav Saboo}
\authornotemark[1]
\email{raghav.saboo@doordash.com}
\affiliation{%
  \institution{DoorDash Inc.}
  \streetaddress{303 2nd Street, Suite 800}
  \city{San Francisco}
  \state{California}
  \postcode{94107}
  \country{USA}
}

\author{Martin Wang}
\email{martin.wang@doordash.com}
\orcid{0009-0004-9002-1880}
\affiliation{%
  \institution{DoorDash Inc.}
  \streetaddress{303 2nd Street, Suite 800}
  \city{San Francisco}
  \state{California}
  \postcode{94107}
  \country{USA}
}

\author{Sudeep Das}
\email{sudeep.das2@doordash.com}
\orcid{0000-0002-1754-5811}
\affiliation{%
  \institution{DoorDash Inc.}
  \streetaddress{303 2nd Street, Suite 800}
  \city{San Francisco}
  \state{California}
  \postcode{94107}
  \country{USA}
}

\begin{abstract} 
In multi-vertical e-commerce platforms like DoorDash, relatively newer product verticals such as grocery and retail present a significant opportunity for personalization innovation. A key challenge lies in solving the "cold start" problem for users. This paper introduces a novel framework for enhancing recommendation quality by transferring knowledge from data-rich verticals (e.g., restaurants at DoorDash) to data-sparse ones. We leverage Large Language Models (LLMs) to perform generative inference, synthesizing sparse, high-dimensional features that encapsulate latent user affinities. Specifically, we employ a hierarchical Retrieval-Augmented Generation (RAG) pipeline to derive multi-level taxonomic features from user restaurant order histories and search queries. These generated features, encoding both long-term cross-vertical preferences and short-term intent, are integrated into a production Multi-Task Learning (MTL) ranking model. We demonstrate through extensive offline and online evaluation that this approach significantly improves personalization and engagement in emerging business verticals, effectively bridging the behavioral data gap.
\end{abstract}

\keywords{Large Language Models, Recommender Systems, Cross-Domain Personalization, Cold-Start Problem, Feature Engineering, Multi-Task Learning, Ranking}

\maketitle

\section{Introduction}
Modern e-commerce ecosystems, characterized by multi-vertical marketplaces like DoorDash, continually expand into new domains such as groceries, retail, and beauty. This expansion introduces a critical personalization challenge: how to provide relevant recommendations in these verticals for users with no engagement history. Concurrently, these platforms possess a wealth of behavioral data from established, high-traffic verticals. This data asymmetry presents a valuable opportunity for strategic knowledge transfer.

This work addresses this challenge by conceptualizing user behavior in established verticals as a source of rich, latent signals that can inform preferences in emerging ones. We posit that large language models are powerful tools for semantic feature engineering \cite{wu2024survey, harrison2023zero, vats2024exploring, wei2024llmrec}. They can distill unstructured data, such as restaurant orders and search queries \cite{ma2021event, sinha2020spir}, into structured, interpretable representations of user affinity.

Our core contribution is a novel methodology that employs a hierarchical Retrieval-Augmented Generation (RAG) framework \cite{huang2025retrieval, wang2025archrag} to infer user affinities at multiple levels of a product taxonomy. This structured inference pipeline mitigates the risk of hallucination and enhances the fidelity of the generated features. By injecting these LLM derived features into our production Multi-Task Learning (MTL) ranking model, we effectively enrich the user representation, enabling the ranker to discern nuanced cross-vertical preferences even in the absence of direct historical data. This approach directly confronts the cold-start problem \cite{zhao2024recommender, wang2025survey, huang2025large} and demonstrates a practical path toward building more holistic and adaptive recommender systems.

\section{System Architecture and Methodology}
Our production recommendation system employs a multi-stage architecture, prominently featuring Two-Tower Embedding (TTE) models for candidate retrieval \cite{huang2020embedding} and a Multi-Task Learning (MTL) model for fine-grained ranking \cite{chapelle2010multi, ma2018modeling, ma2018entire}. This research focuses on augmenting the feature space of the MTL ranker to enhance its ability to model cross-domain preferences. We introduce two novel classes of LLM-synthesized features:

\begin{itemize}[leftmargin=*,label=\textbf{--}]
    \item \textbf{Long-Term Cross-Vertical Affinity Profile:} Captures a consumer's latent, historical preferences for taxonomy categories, inferred from their cumulative restaurant order history.
    \item \textbf{Short-Term Intent Profile:} Models a consumer's recent, transient interests in specific taxonomy categories, as indicated by their on-platform search activity.
\end{itemize}

\subsection{Generative Feature Synthesis via Hierarchical RAG}
\begin{figure*}[htbp]
  \centering
  \includegraphics[width=0.8\textwidth]{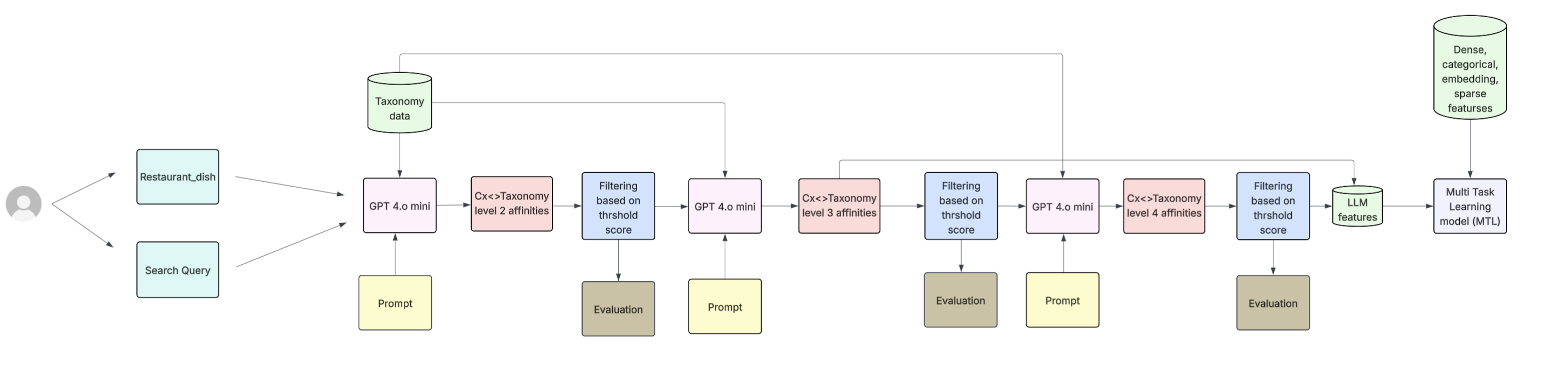}
  \caption{System flowchart illustrating the hierarchical RAG pipeline for generating user affinity features and their integration into the MTL ranking model.}
  \label{fig:flowchart}
\end{figure*}

We employ LLMs to map unstructured user activity (restaurant orders, search queries) to our internal, four-level hierarchical product taxonomy (L1–L4; e.g., Dairy \& Eggs $\rightarrow$ Cheese $\rightarrow$ Hard Cheeses $\rightarrow$ Cheddar). Using a 20\% sample of consumer data from the preceding three months, we execute the hierarchical RAG process depicted in Figure \ref{fig:flowchart}.

The process operates via cascaded inference. First, the model identifies broad, top-level (e.g., L1, L2) taxonomic affinities from the input signals. These initial, high-confidence classifications then constrain the generative search space for subsequent, more granular retrievals at lower taxonomy levels (e.g., L3). This iterative refinement strategy enhances precision and relevance by preventing the model from generating plausible but incorrect subcategories. For our MTL ranker, we concentrate on L2 and L3 affinities, as L1 is often too general and L4 suffers from excessive sparsity.

\subsection{Prompt Engineering and Inference Control}
To structure the model's input, we concatenate historical restaurant names and ordered items in chronological order, prioritizing recent behavior. Search queries are similarly sequenced. This temporal ordering helps the model capture evolving preferences. The prompt is enriched with contextual information, including the target taxonomy structure and anonymized consumer profile attributes.

To ensure deterministic and high-fidelity output, we set the inference temperature to 0.1. Critically, the prompt instructs the model to return a confidence score for each generated affinity and to only output taxonomies exceeding a confidence threshold of 0.8. This acts as a self-correction mechanism, filtering out low-confidence or spurious associations.

Prior to implementing the aforementioned improvements in the prompt, the extracted4 taxonomies from restaurant orders included some irrelevant categories, resulting in poorer alignment with the actual cuisine. For example, as shown in Table~\ref{tab:LLM_prompt_improvements}, a consumer placing orders from an Indian restaurant was previously associated with less relevant affinities, such as "Sandwiches" which do not accurately capture the nuances of Indian cuisine. After applying the improved prompt engineering techniques, the model instead produced more specific and appropriate taxonomies, such as "Specialty Breads (Naan)" thereby significantly enhancing the relevance of the taxonomic associations.

\begin{table*}[htbp]
\centering
\caption{Improvements in the results with prompt engineering techniques}
\label{tab:LLM_prompt_improvements}
\renewcommand{\arraystretch}{1.5} 
\begin{tabularx}{\textwidth}{
    @{} 
    >{\raggedright\arraybackslash}p{0.16\textwidth} | 
    >{\raggedright\arraybackslash}X | 
    >{\raggedright\arraybackslash}X | 
    >{\raggedright\arraybackslash}X 
    @{}
}
\toprule
\textbf{Signal Type} & \textbf{Consumer Input History} & \textbf{Before (LLM Output)}  & \textbf{After (LLM Output)} \\
\midrule
Restaurant Orders & 
Royal Spice: Naan; Butter Chicken; Vegetable Samosa & 
Sandwiches, Burgers \& Wraps, Entrees, Appetizers \& Sides, Chicken &
Specialty Breads, Naan, Vegetable Sides, Chicken \\
\bottomrule
\end{tabularx}
\end{table*}

\begin{tcolorbox}[colback=gray!10, colframe=gray!60, title= Prompt Example]
\texttt{CONTEXT\_SETUP = "You are a recommendation engine. Given a consumer's order history and an allowed L3 taxonomy, infer up to 50 relevant L3 categories. Focus on the cuisine of the restaurant and use only categories present in the provided taxonomy (case-insensitive match, output exact taxonomy spelling). Do not invent categories. Assign a confidence score in [0,1]; include only categories with confidence >= 0.80. Sort by confidence descending; break ties alphabetically. If no category meets the threshold, return an empty list. Output must be strict JSON"}
\\
\texttt{OPERATING\_RULES = "Map items to the most specific applicable L3. If a dish could map to multiple categories, choose the best-supported one. If the result is ambiguous and confidence <0.80, exclude it."}
\\
\texttt{TAXONOMY\_INFO = "The allowed L3 categories are: [L3 taxonomy list]"}
\\
\texttt{USER\_HISTORY = "The consumer has ordered the following dishes from the restaurants in chronological order (store || dish): [restaurant name || dish name], ..."}
\\
\texttt{prompt = CONTEXT\_SETUP + OPERATING\_RULES + " " + USER\_HISTORY + " " + TAXONOMY\_INFO}
\end{tcolorbox}

\subsection{Model Selection and Cost Optimization}
A cost-performance analysis of various models, including GPT-4o and GPT-4o-mini, revealed that GPT-4o-mini delivered comparable output quality for this specific task at a substantially lower computational cost. To further optimize, we implemented a prompt-caching strategy. The static portion of the prompt (containing instructions and taxonomy) is cached, and only the dynamic user history portion is appended for each inference call. This token-level optimization, combined with a just-in-time feature materialization strategy (updating affinities only upon new user actions), reduced overall computational costs by approximately 80\%.

\subsection{Feature Materialization and Serving}
A daily batch pipeline computes and updates the affinity features. These features are materialized to our data lake for model training and simultaneously propagated to a low-latency online feature store for real-time inference during serving.

\subsection{Qualitative Feature Evaluation}
Table \ref{tab:llm_examples} provides illustrative examples of the taxonomic affinities generated. To quantitatively assess the quality of the features, we conducted two studies, one based on human evaluation and the other based on LLM as a judge with the more powerful model GPT-4o, scoring the relevance of personalization on a 3-point scale. As shown in Table \ref{tab:llm-eval-human} for human evaluation and Table \ref{tab:llm-eval-llmjudge} for LLM as a judge evaluation, features derived from search queries exhibit higher personalization scores, which aligns with the explicit nature of search intent compared to the implicit signals from order history. 

\begin{table*}[htbp]
\centering
\caption{Illustrative examples of LLM-generated category recommendations from consumer signals.}
\label{tab:llm_examples}
\renewcommand{\arraystretch}{1.5} 
\begin{tabularx}{\textwidth}{@{} >{\raggedright\arraybackslash}p{0.1\textwidth} | >{\raggedright\arraybackslash}X | >{\raggedright\arraybackslash}X @{}}
\toprule
\textbf{Signal Type} & \textbf{Consumer Input History} & \textbf{Generated L3 Taxonomy Affinities (LLM Output)} \\
\midrule
Restaurant Orders & 
Taco Bell: Cantina Chicken Crispy Taco; Cheese Quesadilla \newline
Royal Spice: Cheese Naan; Butter Chicken \newline
Starbucks: White Chocolate Mocha 
& 
Tacos, Chicken, Cheese, Naan, Specialty Breads, Coffee \\
\hline
Search Queries & Protein bar, drink, pop tart, protein, yogurt, healthy snacks & Cereal \& Granola Bars, Packaged Snacks, Yogurt, Juices \& Smoothies, Protein Supplements, Nutrition Shakes, Energy Drinks, Sour Cream \& Dips \\
\bottomrule
\end{tabularx}
\end{table*}

\begin{table*}[h]
\centering
\caption{Human Evaluation of LLM-Generated Feature Personalization. N=1000 samples per signal.}
\label{tab:llm-eval-human}
\begin{tabular}{lccc}
  \toprule
  \textbf{Signal Source} & \textbf{Not Personalized} & \textbf{Partially Personalized} & \textbf{Highly Personalized} \\
  \midrule
  Restaurant Orders & 17.7\% & 29.3\% & 53.0\% \\
  Search Queries & 6.8\% & 22.5\% & 70.7\% \\
  \bottomrule
\end{tabular}
\end{table*}

\begin{table*}[h]
\centering
\caption{LLM Evaluation of LLM-Generated Feature Personalization (GPT-4o). N=1000 samples per signal.}
\label{tab:llm-eval-llmjudge}
\begin{tabular}{lccc}
  \toprule
  \textbf{Signal Source} & \textbf{Not Personalized} & \textbf{Partially Personalized} & \textbf{Highly Personalized} \\
  \midrule
  Restaurant Orders & 15.6\% & 27.8\% & 56.6\% \\
  Search Queries & 8.2\% & 30.2\% & 61.6\% \\
  \bottomrule
\end{tabular}
\end{table*}

\subsection{Multi-Task Learning (MTL) Architecture}
Our MTL ranker is designed to jointly optimize for multiple objectives (e.g., click-through rate, add-to-cart, purchase). The total loss $\mathcal{L}$ is a weighted sum of individual task losses $\mathcal{L}_t$:
$$
\mathcal{L} = \sum_{t=1}^T \alpha_t \,\mathcal{L}_t\bigl(\hat{y}_t, y_t\bigr),
$$
where $\hat{y}_t$ is the model's prediction for task $t$ and $\alpha_t$ is a task-specific weight. We augment the model's input feature space by concatenating our LLM-generated features with existing feature vectors:
\begin{itemize}[leftmargin=*]
    \item \(\mathbf{u}_{\mathrm{LLM}} \in \mathbb{R}^d\): The new sparse features representing user affinities, derived from both restaurant orders and search queries.
    \item \(\mathbf{u}_{\mathrm{eng}} \in \mathbb{R}^p\): The existing user engagement feature vector (e.g., historical interactions).
    \item \(\mathbf{i}_{\mathrm{eng}} \in \mathbb{R}^q\): The item feature vector (e.g., category, brand, price).
\end{itemize}
The augmented user and item vectors are defined as:
$$
\mathbf{u}_{\mathrm{aug}} = [\mathbf{u}_{\mathrm{eng}}; \mathbf{u}_{\mathrm{LLM}}], 
\quad 
\mathbf{i} = [\mathbf{i}_{\mathrm{eng}}].
$$
Variable-length categorical features, such as the lists of taxonomy IDs in $\mathbf{u}_{\mathrm{LLM}}$, are handled by mapping each ID to a dense vector via a shared embedding table. The embeddings corresponding to a list are then aggregated into a fixed-size representation using mean pooling. This technique efficiently handles jagged input tensors and promotes parameter sharing. The final concatenated feature vector feeds into a shared MLP trunk, $\phi$, followed by task-specific prediction heads:
\begin{equation}
    \mathbf{z} = \phi\bigl([\mathbf{u}_{\mathrm{aug}}, \mathbf{i}]\bigr), \quad
    \hat{y}_t = \sigma\bigl(\mathbf{w}_t^\top \mathbf{z} + b_t\bigr).
\end{equation}
Here, $\sigma$ is an activation function (e.g., sigmoid), and $\mathbf{w}_t, b_t$ are the weights and bias of the prediction head for task $t$.

\section{Experimental Evaluation}
We conducted rigorous set of offline and online experiments to measure the impact of LLM‑generated features on our item‑ranking model trained on three months of data. Offline evaluation was performed on a 15‑day holdout using conversion as the ground‑truth label, and results are reported with the metrics described in the next section.

\subsection{Experimental Setup}
We compare the performance of two models using standard evaluation metrics: Area Under the Receiver Operating Characteristic Curve (AUC-ROC) for classification performance and Mean Reciprocal Rank (MRR) for ranking quality.
\begin{itemize}[leftmargin=*]
    \item \textbf{Baseline Model}: The production MTL item ranking model, trained exclusively on historical user engagement and item attribute features.
    \item \textbf{Proposed Model}: An identical MTL architecture augmented with the LLM-derived user affinity features ($\mathbf{u}_{\mathrm{LLM}}$).
\end{itemize}

\subsection{Offline Evaluation and Cohort Analysis}
\begin{figure}[h]
  \centering
  \includegraphics[width=\linewidth]{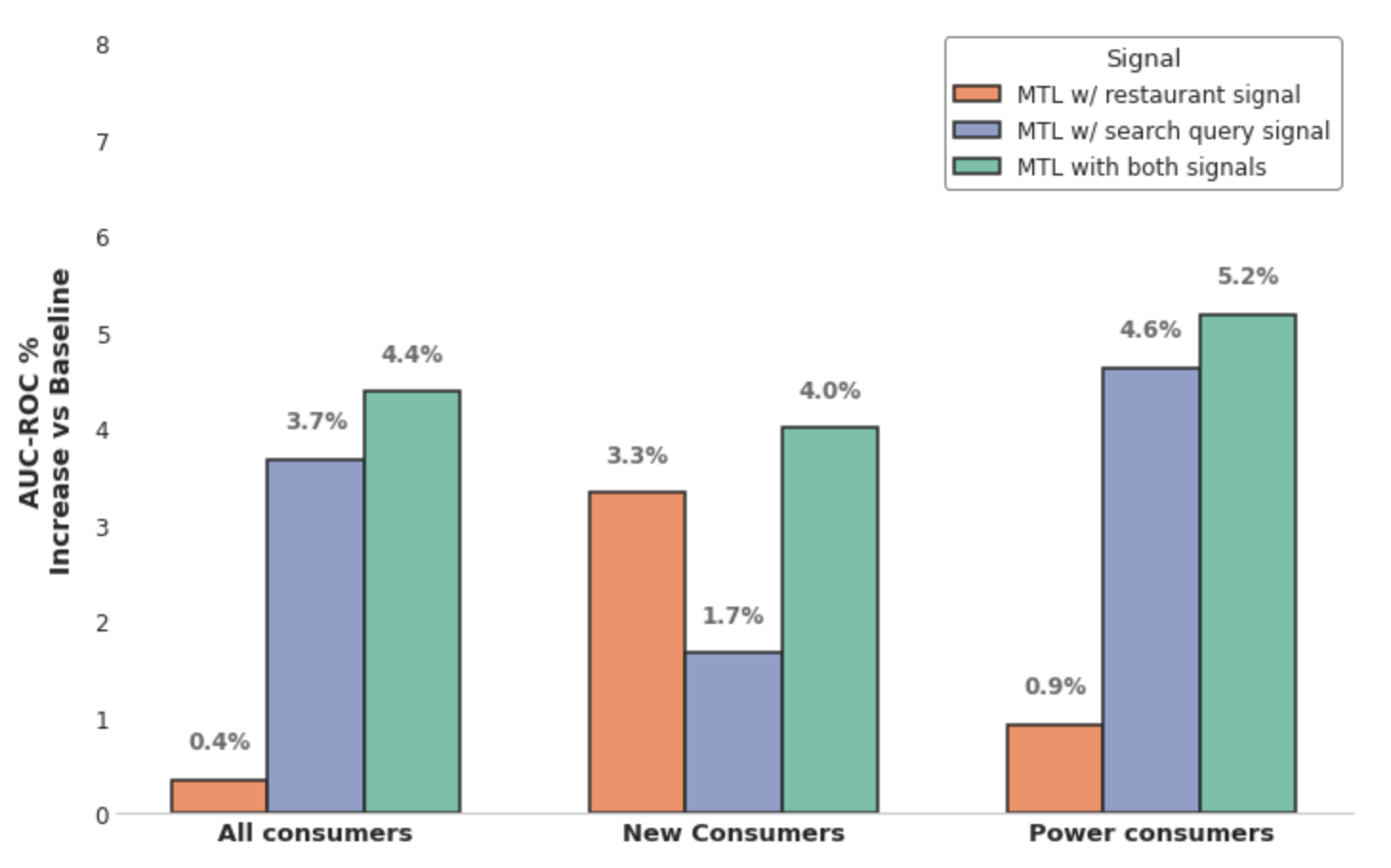}
  \caption{Relative improvement (\%) in AUC-ROC for the Proposed Model over the Baseline across different consumer cohorts.}
  \label{fig:auc_plot}
\end{figure}

\begin{figure}[h]
  \centering
  \includegraphics[width=\linewidth]{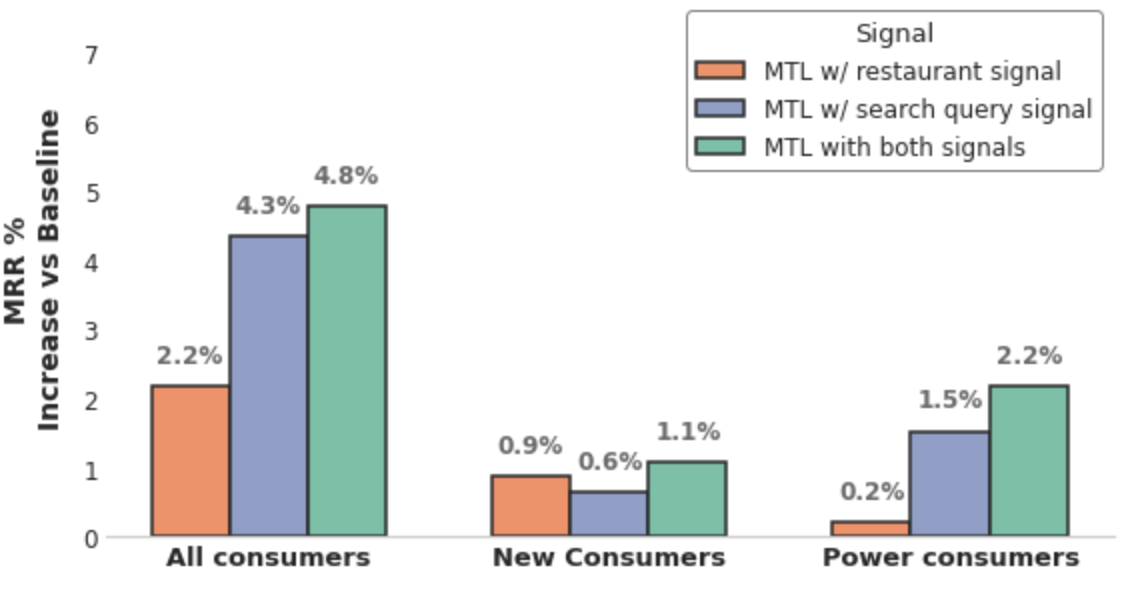}
  \caption{Relative improvement (\%) in MRR for the Proposed Model over the Baseline across different consumer cohorts.}
  \label{fig:mrr_plot}
\end{figure}

We evaluated performance across the general user population and on two specific cohorts critical to our business: "cold-start" consumers (new to non-restaurant verticals) and "power" consumers (highly active in these verticals). The results, summarized in Figures \ref{fig:auc_plot} and \ref{fig:mrr_plot}, demonstrate a consistent and significant performance uplift. 

Key findings include:
\begin{itemize}
    \item \textbf{Overall Population:} The proposed model achieved a 4.4\% relative lift in AUC-ROC and a 4.8\% lift in MRR, indicating a broad improvement in ranking quality.
    \item \textbf{Cold-Start Consumers:} This cohort benefited most from the restaurant order signal, with the combined signals yielding a 4.0\% lift in AUC-ROC and a 1.1\% lift in MRR. This validates our hypothesis that historical taste preferences can be effectively transferred across verticals.
    \item \textbf{Power Consumers:} This group saw the largest gains from the search query signal, which captures short-term intent. The model achieved a 5.2\% lift in AUC-ROC and a 2.2\% lift in MRR, showcasing its ability to adapt to recent user needs.
\end{itemize}

\subsection{Online Shadow Deployment}
\begin{figure}[h]
  \centering
  \includegraphics[width=\linewidth]{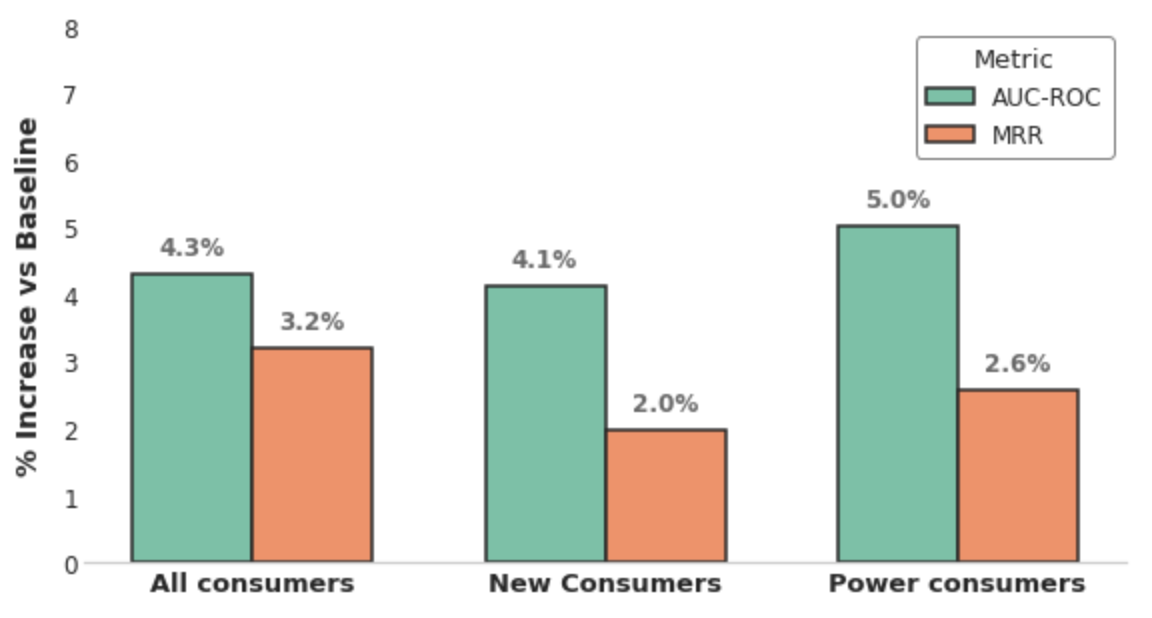}
  \caption{Relative improvement (\%) in online shadow traffic metrics for the Proposed Model versus the Baseline.}
  \label{fig:online_plot}
\end{figure}

To validate our offline findings in a production environment, we conducted an online evaluation by shadowing live traffic. The results, shown in Figure \ref{fig:online_plot}, were consistent with the offline analysis. For the general population, the LLM-augmented model demonstrated a 4.3\% improvement in AUC-ROC and a 3.2\% increase in MRR over the baseline, confirming the real-world efficacy of our approach.

\section{Conclusion and Future Work}
This work successfully demonstrates the efficacy of leveraging LLMs as a semantic bridge to transfer knowledge from data-rich to data-sparse domains within a multi-vertical marketplace. By employing a hierarchical RAG framework to synthesize high-fidelity user affinity features, we significantly enhanced the performance of our production MTL ranking model, particularly for cold-start users. 

Our future research agenda will proceed along several promising vectors. First, we will investigate more advanced prompting paradigms (e.g., Chain-of-Thought, self-correction) and explore domain-specific fine-tuning of smaller, more efficient LLMs to further improve feature quality and reduce costs. Second, we plan to integrate these generative features earlier in the recommendation funnel, specifically within the candidate retrieval stage, to create a more synergistic, end-to-end personalized system. Finally, we will explore the temporal dynamics of these affinities to build more adaptive, session-aware recommendation models.

\section{Author Bio}
Nimesh Sinha is a Senior Machine Learning Engineer at DoorDash, where he specializes in personalizing consumer experiences for the company’s new verticals business. Previously, he built advanced search and personalization models for Walmart’s e-commerce platform. Nimesh has also contributed to machine learning initiatives at Bird and Barnes \& Noble Education. He holds a master’s degree in Data Science from the University of San Francisco and an integrated master’s in Applied Physics from IIT-ISM Dhanbad.

\bibliographystyle{ACM-Reference-Format}
\balance
\bibliography{references}

\end{document}